\documentclass[aps,prx,reprint,superscriptaddress]{revtex4-2}

\usepackage{amsmath}
\usepackage{amssymb}
\usepackage{svg}
\usepackage[separate-uncertainty=true]{siunitx}
\definecolor{red}{RGB}{255,20,35}
\definecolor{blue}{RGB}{25,102,255}
\usepackage{xcolor}
\usepackage[normalem]{ulem}
\newcommand\mkst{\bgroup\markoverwith
{\textcolor{red}{\rule[.5ex]{2pt}{0.4pt}}}\ULon}
\newcommand\cwst{\bgroup\markoverwith
{\textcolor{blue}{\rule[.5ex]{2pt}{0.4pt}}}\ULon}

\usepackage{hyperref}
\usepackage{physics}
\usepackage{upgreek} 
\usepackage[T1]{fontenc}
\usepackage{booktabs}
\usepackage{float}

\hypersetup{%
	colorlinks,
	linkcolor={blue}, %
	citecolor={red}, %
	urlcolor={red} %
}

\def\RLEaffil{Research Laboratory of Electronics, Massachusetts Institute of Technology, Cambridge, MA 02139, USA}
\def\UREGaffil{Institute of Experimental and Applied Physics and Halle-Berlin-Regensburg Cluster of Excellence CCE, University of Regensburg, 93040 Regensburg, Germany}
\def\Physaffil{Department of Physics, Massachusetts Institute of Technology, Cambridge, MA 02139, USA}
\def\EECSaffil{Department of Electrical Engineering and Computer Science, Massachusetts Institute of Technology, Cambridge, MA 02139, USA}
\def\LLaffil{Lincoln Laboratory, Massachusetts Institute of Technology, Lexington, MA 02421, USA}

\def\Qolaffil{Qolab, Madison, Wisconsin 53706, USA}
\def\UWaffil{Department of Physics, University of Wisconsin-Madison, Madison, Wisconsin 53706, USA}
\def\DecayRate{\mathit{\Gamma}_1}

\begin{document}
	
	
	\title{Adaptive Spectroscopy of Fast Two-level-system Dynamics in Superconducting Qubits}
	
\author{Fabrizio~Berritta} 
\email{fabrizio.berritta@mit.edu}
\affiliation{\RLEaffil}
\affiliation{\UREGaffil}
\author{David~Pahl}
\affiliation{\RLEaffil}
\affiliation{\EECSaffil}
\author{Lukas~Pahl}
\affiliation{\RLEaffil}
\affiliation{\EECSaffil}
\author{William~P.~Banner}
\affiliation{\RLEaffil}
\author{Gabriel~Cutter}
\affiliation{\RLEaffil}
\affiliation{\EECSaffil}
\author{Jan~A.~Krzywda}
\affiliation{Lorentz Institute for Theoretical Physics \& Leiden Institute of Advanced Computer Science, Universiteit Leiden, 2311 EZ Leiden, The Netherlands}
\author{Spencer~Weeden}
\affiliation{\UWaffil}
\author{Shravan~Patel}
\affiliation{\UWaffil}
\author{Paul~Buttles}
\affiliation{\Qolaffil}
\author{Stanislav~Eilhart}
\affiliation{\Qolaffil}
\author{Michael~Gingras}
\affiliation{\LLaffil}
\author{Bethany~M.~Niedzielski}
\affiliation{\LLaffil}
\author{Robert~McDermott}
\affiliation{\UWaffil}
\affiliation{\Qolaffil}
\author{Mollie~E.~Schwartz}
\affiliation{\LLaffil}
\author{Kyle~Serniak}
\affiliation{\RLEaffil}
\affiliation{\LLaffil}
\author{Max~Hays}
\affiliation{\RLEaffil}
\author{Jeffrey~A.~Grover}
\affiliation{\RLEaffil}
\author{William~D.~Oliver}
\email{william.oliver@mit.edu}
\affiliation{\RLEaffil}
\affiliation{\EECSaffil}
\affiliation{\Physaffil}

\date{August 3, 2026}
\begin{abstract}
Parasitic two-level-system (TLS) defects are a major source of energy relaxation and temporal instability in superconducting quantum processors. Our sub-second adaptive spectroscopy reveals telegraphic switching of TLSs with a characteristic timescale of a few seconds and spectral diffusion with diffusivity $D\approx\SI{0.9}{\mega\hertz\squared\per\second}$. These timescales are about $3\times 10^2$ times faster than what is observed in conventional nonadaptive spectroscopy, which typically requires hours of measurement time. We resolve such fast dynamics on a field-programmable gate array (FPGA)-based controller that enables measurement of frequency- and time-resolved relaxations with sub-second temporal resolution in flux-tunable superconducting qubits. We observe similar defect dynamics across multiple qubits in independently fabricated devices measured in different laboratories. We correlate TLS-induced fluctuations with gate-level errors using randomized benchmarking. Our results reveal a previously inaccessible regime of frequency-resolved TLS dynamics and redefine the timescales relevant to TLS-aware characterization and calibration of superconducting quantum processors.

\end{abstract}
	\maketitle
	
	\section{Introduction} 

Two-level system (TLS) defects are a main source of decoherence in superconducting qubits~\cite{muller2019, Siddiqi2021,Murray2021} and semiconductor spin qubits~\cite{rojas2023spatial,ye2024, donnelly2025noise, Rojas2026}, although their detailed microscopic origins remain unknown. An individual defect can cause discrete telegraphic jumps in the qubit frequency when dispersively coupled or modulate the qubit energy-decay rate $\DecayRate$ when nearly resonant with the qubit energy splitting~\cite{lisenfeld2016decoherence,Schloer2019}. Ensembles of TLSs can also produce approximately $1/f$ charge or flux noise from sub-Hz to kHz frequencies. These fluctuations degrade the fidelity of quantum operations and constitute non-Markovian noise~\cite{gao2026non} that can compromise quantum error correction~\cite{kam2024detrimental} and increase the cost of error mitigation~\cite{hakoshima2021relationship}.
As the size of quantum processing units (QPUs) increases, the probability that at least one qubit is limited by a TLS grows exponentially. Fast, time-dependent fluctuations make it nontrivial to identify which qubits limit the QPU performance at any given moment.
Moreover, individual TLSs can fluctuate in frequency within a cooldown and rearrange across different cooldowns, even though the total defect count remains approximately unchanged ~\cite{Shalibo2010, Zanuz2024}. 

TLSs distributed across the GHz spectrum, including defects in aluminum oxide and other insulating materials, can interact strongly with superconducting qubits through electric-dipole coupling~\cite{Shalibo2010,lisenfeld2019}. 
Strategies for reducing TLS-induced errors include improvements in device architecture, materials, and fabrication~\cite{Siddiqi2021,mcrae2021,weeden2025, wolff2026structural, degnan2026reducing}; direct control or stabilization of noise sources~\cite{grabovskij2012,lisenfeld2019,bilmes2020, kim2024error, chen2025, dane2025, ye2025stabilizing}; and real-time feedback~\cite{Berritta2026_bistable}.

Switches in relaxation time by an order of magnitude have been observed over timescales of just milliseconds~\cite{Berritta2026_T1}. In fixed-frequency relaxation tracking, only a small fraction of such rapid $\DecayRate(t)$ fluctuations can be attributed to TLS-like dynamics, because the qubit frequency is fixed and the underlying defect cannot be tracked spectrally. Flux-tunable qubits instead provide the spectral degree of freedom needed to map the frequency dynamics of TLS resonances that cause rapid relaxation-time fluctuations.
Previous works in transmon qubits have reported TLS spectral diffusion on timescales of hours~\cite{Klimov2018, bejanin2021, Carroll2022, weeden2025, roy2026two}. 
Accessing faster regimes of TLS dynamics requires estimation protocols that resolve the dynamics before they are averaged out and can rapidly flag outlier qubits and time-dependent fluctuations~\cite{muller2019, Siddiqi2021, Murray2021, mcrae2021} in large QPUs.
Such information can support efficient characterization and error mitigation~\cite{Leroux2025,Drouet2026} by identifying and avoiding frequencies where TLSs are resonant with the qubits~\cite{klimov2024}.
Beyond device screening, fast spectral maps of $\DecayRate$ can also help reveal correlations among interacting defects and quantify the coupling of individual TLSs to fluctuating defect baths, thus allowing rapid experimental tests of microscopic models for TLS dynamics.

Here we extend the adaptive decay-rate estimation method of Ref.~\cite{Berritta2026_T1}, demonstrated in fixed-frequency transmon qubits, to frequency-resolved TLS spectroscopy with flux-tunable transmons. This work is organized around three parts: (i) we describe how FPGA-based adaptive estimates of $\DecayRate$ are converted into time-resolved TLS maps, (ii) we quantify the seconds-scale telegraphic and diffusive motion revealed by these maps, and (iii) we interleave the same spectroscopy with randomized benchmarking to test whether the tracked relaxation-rate fluctuations are visible at the gate-error level.

\section{Protocol} 
\begin{figure}
\centering
\includegraphics{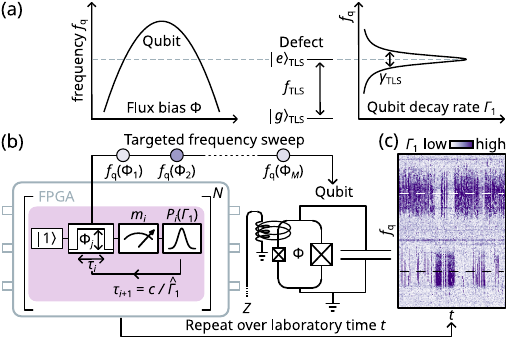}
\caption{\textbf{Adaptive relaxation spectroscopy of TLS-induced loss.} 
(a) Schematic of a flux-tunable qubit swept through resonance with a two-level-system defect. When $f_\mathrm{q}\approx f_{\mathrm{TLS}}$, the defect causes a peak in the qubit decay rate $\DecayRate$, whose spectral width $\gamma_{\mathrm{TLS}}$ is set by the defect linewidth and its coupling to the qubit.
(b) Adaptive relaxation spectroscopy schedule. At each of the $M$ target qubit frequencies, the FPGA estimates the qubit decay rate $\DecayRate$ (purple box) over $N$ adaptive probe cycles. During cycle $i$, the controller prepares the transmon in $\ket{1}$ and selects the wait time $\tau_i$ from the current Bayesian probability distribution ${\cal P}_i(\DecayRate)$. During $\tau_i$ the qubit is tuned ($\Phi_j$) to the target frequency. Afterwards, the controller measures the binary outcome $m_i$, and updates ${\cal P}_{i+1}(\DecayRate)$. 
(c) Example time-resolved TLS map $\DecayRate(f_{\mathrm{q}},t)$ measured by repeating the frequency sweep over laboratory time $t$. The decay-rate peak (dark purple) switches between two discrete frequencies (dashed lines), resulting in the characteristic signature of a telegraphic TLS.
}
\label{fig:fig1}
\end{figure}

We implement the adaptive relaxation-spectroscopy protocol on two superconducting-qubit devices, Device A and Device B, containing transmons with DC SQUIDs~\cite{Krantz2019,Blais2021}. Device A consists of uncoupled flux-tunable transmons, whereas Device B has flux-tunable transmons connected by tunable couplers. The devices are fabricated with different processes and are measured in two different laboratories: Device A at the Israeli Quantum Computing Center (IQCC) and Device B at MIT. The qubits have maximum frequencies near $\SI{5.1}{\giga\hertz}$ in Device A and $\SI{6.2}{\giga\hertz}$ in Device B. A commercial controller with integrated FPGA hardware (Quantum Machines OPX1000) synthesizes the high-frequency waveforms applied to the on-chip $Z$ and $XY$ control lines and demodulates the resonator readout signals. The Supplemental Material provides further details on the experimental setups and device fabrication~\cite{supplementary}.

Flux-biased TLS spectroscopy estimates the decay rate $\DecayRate$ while tuning the qubit frequency $f_{\mathrm{q}}$ with the external flux $\Phi$ [see Fig.~\ref{fig:fig1}(a)]. When the qubit is resonant with the defect, energy is transferred to the TLS, which subsequently relaxes into its local environment, giving rise to an enhanced qubit decay rate. 
Bayesian parameter estimation provides a natural strategy for low-latency adaptive estimation and control~\cite{Gebhart2023,Kurchin2024,Arshad2024, Berritta2024a, Berritta2024b, Park2025,Berritta2025_FBS,Berritta2026_T1, belliardo2026}. Rather than storing only a point estimate, the controller keeps track of a probability distribution for $\DecayRate$ and uses it to choose the next probe wait time after each single-shot measurement. Here we use the protocol of Ref.~\cite{Berritta2026_T1} to estimate $\DecayRate$ more than $100\times$ faster than conventional decay measurements, enabling fast, frequency-resolved spectroscopy of TLSs.

For each applied flux bias $\Phi_j$ ($1\leq j\leq M$), corresponding to a qubit frequency $f_{\mathrm{q}}(\Phi_j)$, the controller estimates $\DecayRate$ from $N$ adaptive probe cycles ($1\leq i\leq N$), as schematized in Fig.~\ref{fig:fig1}(b). 
At the start of probe cycle $i$, a conditional $X_{\pi}$ pulse prepares the qubit in $\ket{1}$ at $\Phi=0$. 
The qubit is then tuned to the target frequency for the adaptive wait time $\tau_i$ (described below), returned to $\Phi=0$, after which dispersive readout returns $m_i\in\{0,1\}$, corresponding to the states $\ket{0}$ and $\ket{1}$. 
Each single-shot outcome updates the probability distribution for $\DecayRate$.

\begin{figure*}
    \centering
    \includegraphics{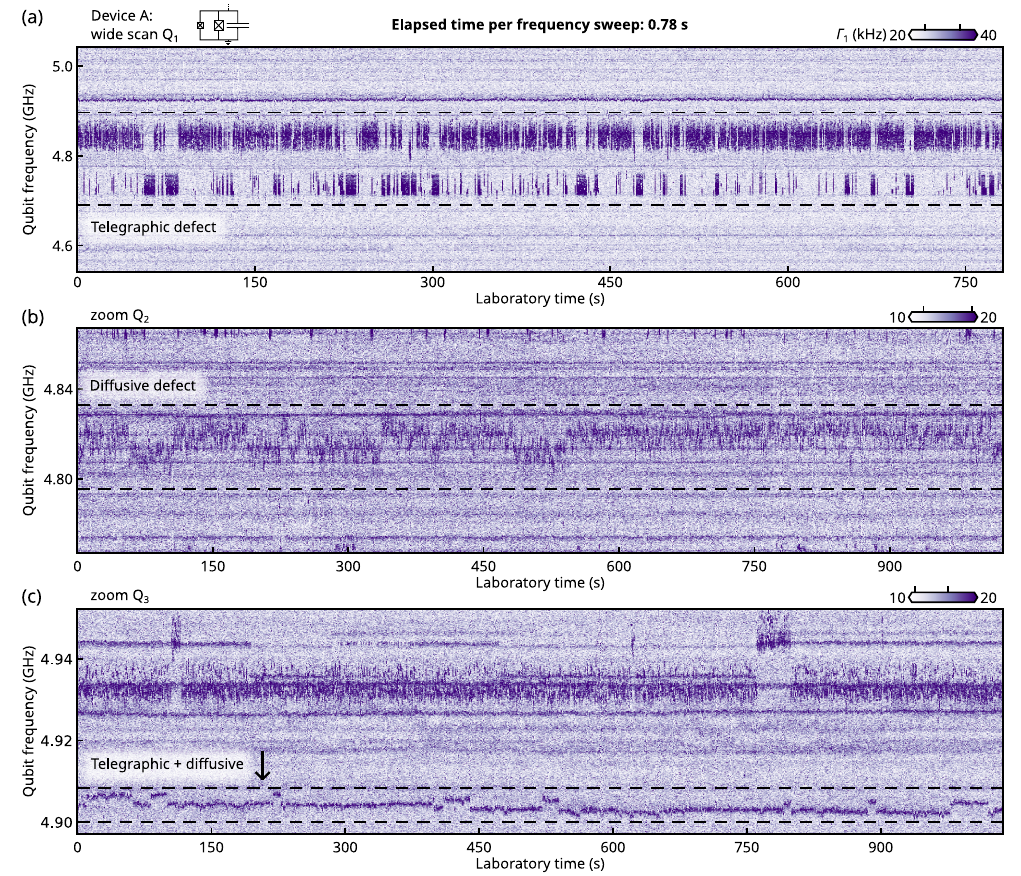}
    \caption{\textbf{Fast, time-resolved spectroscopy of TLS-induced relaxation.} 
Adaptive estimates of the qubit decay rate $\DecayRate$ as a function of qubit frequency and laboratory time reveal spectrally and temporally fluctuating TLSs in Device A on timescales of seconds. The device comprises uncoupled flux-tunable transmon qubits.
(a) Wide frequency scan of $Q_1$, showing multiple decay-rate peaks, including a telegraphic defect with few-second switching highlighted by dashed lines. A Bayesian analysis based on a two-state hidden Markov model yields
a characteristic switching correlation time $\tau_{\mathrm{sw}}\approxeq\SI{2.2}{\second}$. Each column in this plot requires an average estimation time of only approximately $\SI{0.78}{\second}$ spanning 500 qubit frequencies. 
(b) Zoomed scan of $Q_2$, showing a diffusive TLS feature with fitted diffusivity $D\approx\SI{0.93}{\mega\hertz\squared\per\second}$. 
(c) Zoomed scan of $Q_3$. The arrow highlights a TLS with telegraphic switching on top of a diffusive frequency drift.
}
\label{fig:fig2}
\end{figure*}

In the quasistatic approximation, Bayes' rule gives
$
{\cal P}_{i+1}(\DecayRate)
\propto
{\cal P}_{i}(\DecayRate)
P(m_{i+1}|\DecayRate,\tau_{i+1}),
$
where the prior ${\cal P}_{i}(\DecayRate)$ is the probability distribution for $\DecayRate$ after the $i^{\mathrm{th}}$ probe cycle and therefore depends on all previous wait times and measurement outcomes. 
The likelihood $P(m_{i+1}|\DecayRate,\tau_{i+1})$ is the probability of measuring outcome $m_{i+1}$ after initialization and waiting for $\tau_{i+1}$~\footnote{The likelihood function is
$
P(m|\DecayRate,\tau)=1-m-(-1)^m[\beta+(1-\alpha-\beta)e^{-\DecayRate\tau}],
$
where $\alpha$ and $\beta$ are the misclassification probabilities for measuring $\ket{0}$ when the true state at the beginning of the measurement is $\ket{1}$ and measuring $\ket{1}$ when the true state is $\ket{0}$, respectively.}, and the posterior ${\cal P}_{i+1}(\DecayRate)$ is the updated distribution after the subsequent cycle. Operationally, the key idea is that the probability distribution is stored and updated on the controller using only two parameters, as detailed in Ref.~\cite{Berritta2026_T1}. 
The updated estimate $\hat{\DecayRate}$ sets the next wait time, $\tau_{i+1}=c/\hat{\DecayRate}$, with $c=0.1$. 
After each final $\hat{\DecayRate}$ estimate, the prior is reinitialized so that estimates at different flux points and laboratory times are independent. 
Repeating this adaptive estimate over qubit frequency and laboratory time yields the telegraphic TLS map in Fig.~\ref{fig:fig1}(c), which is a zoom-in of the data in Fig.~\ref{fig:fig2}(a). 

\section{Results} \label{sec:results}

\subsection{Fast time-resolved spectroscopy of defect-induced relaxation}

We first apply our protocol to Device A to probe TLSs as a function of qubit frequency with previously unattainable temporal resolution. Figure~\ref{fig:fig2} presents the main TLS maps: Low-latency spectroscopy reveals TLS-induced decay-rate peaks that switch and diffuse on timescales of seconds.

Figure~\ref{fig:fig2}(a) shows the controller estimates $\hat{\DecayRate}$ in $Q_1$. The controller estimates $\DecayRate$ using $N=30$ probe cycles~\footnote{A probe cycle consists of a $\SI{1.5}{\micro\second}$ readout, an approximately $\SI{3}{\micro\second}$ resonator-depletion wait, and the adaptive wait time. The two-parameter posterior approximation keeps the Bayesian update time to about $\SI{2.2}{\micro\second}$} at each value of $\Phi$, sweeping the flux so that $f_\text{Q1}$ spans a $\SI{500}{\mega\hertz}$ detuning range from the upper degeneracy point in $\SI{1}{\mega\hertz}$ steps. Each $\DecayRate$ estimate requires only about $\SI{1.5}{\milli\second}$ of laboratory time. Every frequency sweep is interleaved with a Ramsey experiment at $\Phi = 0$ to verify afterwards that the qubit frequency is calibrated within $\pm\SI{1}{\mega\hertz}$. 

The most striking feature in Fig.~\ref{fig:fig2}(a) is the pair of anticorrelated decay-rate peaks highlighted by the dashed lines. Their anticorrelation suggests that a single resonant TLS switches between two discrete transition frequencies through interactions with another TLS, producing telegraphic fluctuations of the qubit relaxation rate~\cite{Klimov2018}.
From a Bayesian analysis based on a two-state hidden Markov model (see Supplemental Material~\cite{supplementary}), we obtain the
characteristic switching correlation time $\tau_{\mathrm{sw}}=2.20^{+0.24}_{-0.22}\,\si{\second}$, where the quoted value is the posterior median and the uncertainty is
the 68\% credible interval. In contrast to fixed-frequency tracking~\cite{Berritta2026_T1}, flux tunability lets us identify the spectral motion associated with these relaxation events.

The rapid switching events in Fig.~\ref{fig:fig2}(a) are spectroscopically resolved only because each frequency sweep is acquired with a sub-second sampling period. By comparison, previous transmon TLS and coherence-tracking experiments~\cite{Klimov2018,Schloer2019, Carroll2022} typically used sampling intervals of tens of minutes, approximately two orders of magnitude longer than ours, making such second-scale TLS dynamics not directly observable.
Additional horizontal peaks are shown in $\DecayRate$ of Fig.~\ref{fig:fig2}(a), which may be associated with other TLSs or modes in the setup. 
Thus, while slow drifts still require long measurements, our protocol makes it possible to acquire two-dimensional spectroscopy maps with sufficient statistics to characterize fast TLS fluctuations within minutes.

Panels~\ref{fig:fig2}(b) and (c) show zoomed scans from two other qubits on the same device, acquired with similar settings (wider scans are included in the Supplemental Material~\cite{supplementary}). In panel~(b), the highlighted TLS feature undergoes a diffusive process~\cite{Klimov2018}. We extract its center frequency $f_{\mathrm{TLS}}(t)$, compute the mean-squared displacement $\mathrm{MSD}(\tau)=\langle[f_{\mathrm{TLS}}(t+\tau)-f_{\mathrm{TLS}}(t)]^2\rangle_t$, and fit the result to an Ornstein-Uhlenbeck diffusion model. In the short-lag limit, this model gives $\mathrm{MSD}(\tau)\approx 2D\tau$, where $D$ is the diffusivity, and $\tau$ is the lag time. The diffusivity sets the rate at which the TLS frequency wanders. From the fit we obtain $D=0.93^{+0.18}_{-0.22}\,\si{\mega\hertz\squared\per\second}$, where the uncertainty is the 68\% interval from a block bootstrap resampling of the tracked trajectory (see the Supplemental Material~\cite{supplementary}). After converting previously reported diffusivities to the same short-time MSD convention, this diffusivity is about $3\times 10^2$ times larger than values reported from conventional nonadaptive spectroscopy~\cite{Klimov2018, weeden2025}. Panel~(c) illustrates a mixed regime, where slow, approximately diffusive drift is superimposed on fast telegraphic switching. Additional telegraphic jumps are also visible at higher frequencies in panels~(b) and~(c). 

\begin{figure}
    \centering
    \includegraphics{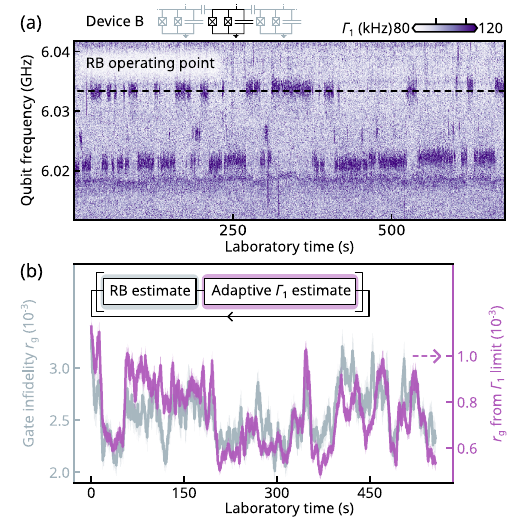}
    \caption{\textbf{TLS-correlated gate-error fluctuations.} (a) Time-resolved TLS map near a telegraphic TLS in Device B, which has coupled flux-tunable transmon qubits. The marked operating point (dashed line) is used for the subsequent randomized-benchmarking measurement in panel~(b). 
    (b) Randomized-benchmarking sequences are interleaved with adaptive $\DecayRate$ estimation at the same operating point. The measured gate infidelity (gray, left axis) qualitatively follows the infidelity expected from the independently estimated relaxation rate (purple, right axis), supporting a correlation between defect-induced $\DecayRate$ fluctuations and gate-level errors. 
    The zero-lag correlation is $\rho_0=0.618$ at the on-TLS operating point, compared with $\rho_0=-0.0507$ in a reference measurement performed a few megahertz away from the TLS resonance~\cite{supplementary}.
}
\label{fig:fig3}
\end{figure}

To test whether these observations are reproducible across independently fabricated devices, we perform the same adaptive spectroscopy on a qubit from Device B measured in a different laboratory.
The result, shown in Fig.~\ref{fig:fig3}(a), is qualitatively similar in terms of timescales to Fig.~\ref{fig:fig2}: a telegraphic TLS with (three) stable frequency configurations superimposed on a slow diffusion. A wider scan is provided in the Supplemental Material~\cite{supplementary}. Because Device A and Device B differ in circuit architecture and operating frequency, we use Device B to demonstrate that fast TLS dynamics are also observed in a distinct device platform, rather than to compare absolute decay-rate values across devices. Together, measurements on Devices A and B show that frequency-resolved TLS motion can be tracked on timescales of seconds.

\subsection{TLS-correlated gate-error fluctuations} 
To assess whether the fast fluctuations of a TLS impact gate performance, we interleave adaptive estimates of $\DecayRate$ at the operating point marked in Fig.~\ref{fig:fig3}(a) with single-qubit randomized benchmarking (RB)~\cite{Knill2008} to estimate the native gate infidelity $r_{\text{g}}$. As described in the inset of Fig.~\ref{fig:fig3}(b), each RB estimate of the average gate infidelity $r_{\text{g}}$ is interleaved with a $\DecayRate$ estimate. Each $r_{\text{g}}$ estimate is obtained from 300 random sequences, with circuit depths stepped logarithmically in powers of two up to 2048, while each $\DecayRate$ estimate uses $N=40$ probe cycles. The average sampling time is approximately $\SI{0.186}{\second}$ and is largely dominated by the RB estimation time.

In the main panel of Fig.~\ref{fig:fig3}(b), we plot $\SI{5}{\second}$ moving averages of the gate infidelity extracted offline from RB (gray curve) and of the infidelity expected from relaxation (purple curve), approximated as $t_{\text{gate}}\DecayRate/3$~\cite{Malley2015} with $t_{\text{gate}} \approx \SI{20}{\nano\second}$. The shaded regions represent 68\% confidence intervals for the moving averages, and they are barely visible because of small uncertainties. For the RB curve, the interval is computed from the standard error of the fitted gate infidelities within the averaging window. For the $\DecayRate$-limited curve, the posterior standard deviations of $\DecayRate$ are combined over the same averaging window. 

The RB infidelity follows the relaxation-limited estimate at the level of slow temporal trends, suggesting that the TLS contributes to the observed gate-error fluctuations. 
Quantitatively, the normalized zero-lag correlation between the $\SI{5}{\second}$ moving-averaged RB infidelity and the corresponding relaxation-limited estimate is $\rho_0=0.618$. To verify that this correlation is associated with the tracked TLS rather than a generic background fluctuation, we repeat the experiment while biasing the qubit frequency a few MHz away from the TLS resonance. In this off-TLS setting, the correlation drops to $\rho_0=-0.0507$ (see the Supplemental Material~\cite{supplementary}). We therefore interpret the tracked TLS as a significant, but not exclusive, contributor to the observed gate-error fluctuations. We tentatively attribute the remaining discrepancies to dephasing-related fluctuations and other RB-sensitive noise sources not captured by $\DecayRate$ alone.

\section{Discussion and outlook} 

The main result of this work is the observation of spectral diffusion of TLSs on timescales of seconds. For the telegraphic feature in Fig.~\ref{fig:fig2}(a), the Bayesian hidden Markov model analysis gives a nominal switching time $\tau_{\mathrm{sw}}\approx\SI{2.2}{\second}$. For the diffusive feature in Fig.~\ref{fig:fig2}(b), the spectral-diffusion analysis gives $D\approx\SI{0.93}{\mega\hertz\squared\per\second}$, about $3\times 10^2$ times larger than previously reported by conventional TLS spectroscopy~\cite{Klimov2018, weeden2025}. 
Our findings are enabled by adaptive Bayesian estimation in flux-tunable transmon qubits: Frequency sweeps combined with low-latency FPGA updates produce time-resolved TLS maps with sub-second sampling over a wide frequency range. 

Interleaved randomized benchmarking shows that the observed TLS dynamics are correlated with fluctuations of gate performance. Because the $\DecayRate$ enhancement correlated with the TLS is estimated online on the controller and requires only a runtime on the order of tens of $1/\DecayRate$, our protocol could be integrated into adaptive control loops with low-latency for error mitigation. For example, spectroscopy could be interleaved with circuit execution and used to temporarily pause operation when the estimated relaxation rate is above a user-defined threshold. 

Our approach is therefore useful for rapid qubit characterization, device-level process feedback, and calibration strategies that account for frequency- and time-dependent TLS fluctuations. Looking ahead, the same framework can be combined with controlled TLS tuning (electric field or strain)~\cite{grabovskij2012, lisenfeld2019, bilmes2020, kim2024error, chen2025, dane2025}, mapping the positions of TLSs~\cite{lisenfeld2026mapping}, AC Stark modulation~\cite{Carroll2022, chen2025}, and probes of background ionizing radiation~\cite{thorbeck2023}. 

While improvements in design, materials, and fabrication are ongoing and worth pursuing, our results motivate a shift from periodic recalibration every few hours to low-latency calibration loops~\cite{Berritta2025_FBS,marciniak2026}. 
By identifying TLS dynamics as a source of rapid spectral and temporal fluctuations of the decay rate, our work provides a better understanding of microscopic defects to improve QPU performance.

\section{Acknowledgments}
We gratefully acknowledge the Superconducting Quantum Information Device Lab (SQuID Lab), led by Morten Kjaergaard at the Niels Bohr Institute, and Ferdinand Kuemmeth of the University of Regensburg for helpful discussions during the early stages of this work. We also thank Tom Dvir for help with the experimental setup at the IQCC. This work was supported by the U.S. Army Research Laboratory and the U.S. Army Research Office under Grant No. W911NF-23-1-0255; the ARO Multi-University Research Initiative under Grant No. W911NF-18-1-0218; IARPA and the Army Research Office under the Entangled Logical Qubits program through Cooperative Agreement No. W911NF-23-2-0212; the U.S. Air Force under Contract No. FA8702-15-D-0001; and the European Union's Horizon Europe research and innovation programme through the Marie Sk{\l}odowska-Curie Actions (MSCA) Postdoctoral Fellowships under Grant Agreement No. 101204890 (HORIZON-MSCA-2024-PF-01). Device A was fabricated at the UW-Madison Wisconsin Center for Nanoscale Technology (wcnt.wisc.edu). The Center is partially supported by the Wisconsin Materials Research Science and Engineering Center (NSF DMR-2309000) and the University of Wisconsin-Madison. J.A.K. acknowledges funding from the Dutch National Growth Fund (NGF) as part of the Quantum Delta NL programme, as well as insightful discussions with Evert van Nieuwenburg. 
Any opinions, findings, conclusions, or recommendations expressed in this material are those of the author(s) and should not be interpreted as necessarily representing the official policies or endorsements of the U.S. Government, the European Union, or the granting authority. Neither the European Union nor the granting authority can be held responsible for them. 

\section{Author contributions}
F.B. conceived the experiment, led the measurements and data analysis, and wrote the manuscript with input from all authors. F.B., D.P., L.P., M.H., J.A.G., and W.D.O.  performed the experiment. J.A.K. provided theoretical contributions. W.P.B., G.C., and P.B. developed experimental infrastructure. S.W. and S.P. designed device A, which was fabricated by S.E. under the supervision of R.M. D.P. and L.P. designed device B, which was fabricated at Lincoln Laboratory by M.G. and B.M.N. under the supervision of M.E.S. and K.S. The project was supervised by M.H., J.A.G., and W.D.O.

	\bibliography{my_bibliography}
 
\end{document}


\beginsupplement


	
\title{Supplemental Material for ``Adaptive Spectroscopy of Fast Two-level-system Dynamics in Superconducting Qubits''}

\author{Fabrizio~Berritta} 
\email{fabrizio.berritta@mit.edu}
\affiliation{\RLEaffil}
\affiliation{\UREGaffil}
\author{David~Pahl}
\affiliation{\RLEaffil}
\affiliation{\EECSaffil}
\author{Lukas~Pahl}
\affiliation{\RLEaffil}
\affiliation{\EECSaffil}
\author{William~P.~Banner}
\affiliation{\RLEaffil}
\author{Gabriel~Cutter}
\affiliation{\RLEaffil}
\affiliation{\EECSaffil}
\author{Jan~A.~Krzywda}
\affiliation{Lorentz Institute for Theoretical Physics \& Leiden Institute of Advanced Computer Science, Universiteit Leiden, 2311 EZ Leiden, The Netherlands}
\author{Spencer~Weeden}
\affiliation{\UWaffil}
\author{Shravan~Patel}
\affiliation{\UWaffil}
\author{Paul~Buttles}
\affiliation{\Qolaffil}
\author{Stanislav~Eilhart}
\affiliation{\Qolaffil}
\author{Michael~Gingras}
\affiliation{\LLaffil}
\author{Bethany~M.~Niedzielski}
\affiliation{\LLaffil}
\author{Robert~McDermott}
\affiliation{\UWaffil}
\affiliation{\Qolaffil}
\author{Mollie~E.~Schwartz}
\affiliation{\LLaffil}
\author{Kyle~Serniak}
\affiliation{\RLEaffil}
\affiliation{\LLaffil}
\author{Max~Hays}
\affiliation{\RLEaffil}
\author{Jeffrey~A.~Grover}
\affiliation{\RLEaffil}
\author{William~D.~Oliver}
\email{william.oliver@mit.edu}
\affiliation{\RLEaffil}
\affiliation{\EECSaffil}
\affiliation{\Physaffil}
	
\date{August 3, 2026}
\maketitle
\tableofcontents


\section{Experimental setups}

\begin{figure*}
    \centering
    \includegraphics{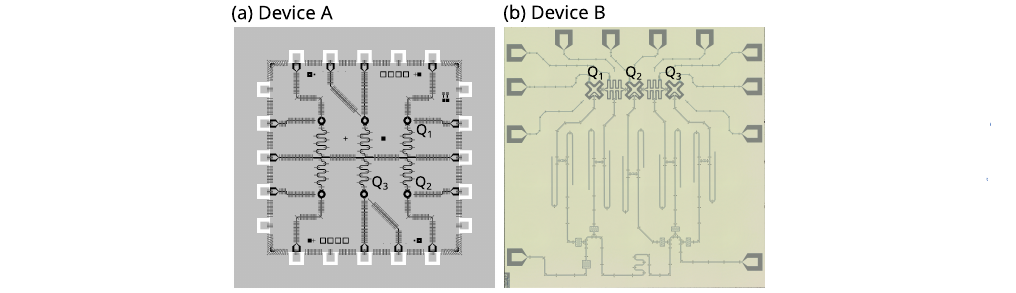}
    \caption{\textbf{Devices A and B}.
    (a) Design layout of Device A, which has six uncoupled flux-tunable transmon qubits and the chip size is $\SI{10}{\milli\meter}\times\SI{10}{\milli\meter}$.
    (b) Optical micrograph of a device nominally identical to Device B. It is a $\SI{5}{\milli\meter}\times\SI{5}{\milli\meter}$ three-qubit chip with flux-tunable transmons connected by tunable couplers. The main text reports measurements from the middle qubit (Q$_2$).}
    \label{fig:SuppFigDevices}
\end{figure*}

Device A is operated at the Israeli Quantum Computing Center (IQCC), while Device B is operated in an academic laboratory at MIT. Figure~\ref{fig:SuppFigDevices}(a,b) shows the layout of Device A and the optical micrograph of a device nominally identical to Device B.
\subsection{Experimental setup at the IQCC and Device A}

The measurements are performed in a Bluefors XLD1000 dilution refrigerator with a base temperature below $\SI{10}{\milli\kelvin}$. A Quantum Machines OPX1000 is used for the readout signal, XY and Z control of the qubit. The average anharmonicity of the qubits is approximately $\SI{-300}{\mega\hertz}$. The OPX1000 includes real-time classical processing with fast analog feedback programmed in QUA software. The readout tone and control lines are attenuated in the cryostat to remove excess thermal photons from higher-temperature stages, and filtered at the mixing chamber. The Z control lines are passively attenuated by $\SI{10}{\decibel}$ at room temperature and $\SI{20}{\decibel}$ inside the cryostat, then filtered using a Quantum Microwave QMC-CRYOIRF-001 low cut-off IR filter. The XY control lines and readout tone, approximately $\SI{6.3}{\giga\hertz}$, are passively attenuated by $\SI{60}{\decibel}$ inside the cryostat and filtered with an RLC Electronics F-30-8000-R low pass filter and Quantum Microwave QMC-CRYOIRF-003 high cut-off IR filter. The transmitted signal from the feedline goes through a high-cutoff IR filter, a Keenlion 4-8~$\SI{}{\giga\hertz}$ bandpass filter, and Low Noise Factory 4-8~$\SI{}{\giga\hertz}$ single and double junction isolators to prevent back reflections from the amplification chain. A high-electron mobility transistor amplifier (LNF-LNA-4-8G) thermally anchored at the $\SI{3}{\kelvin}$ stage amplifies the readout signal. At room temperature, the readout line is again amplified (Narda-MITEQ LNA-40-04000800-07-10P). The device sample is aluminum wirebonded into a custom-made sample mount using a high density of package-to-chip and chip-to-chip wirebonds to ensure proper grounding. The sample mount comprises a printed circuit board and superconducting aluminum enclosure designed to suppress stray microwave and infrared photons. The sample mount is placed inside a light-tight cryoperm can to further reject stray photons and provide magnetic shielding. Both the sample mount and the magnetic shield can are supplied by Qolab. The tunable transition frequency of our transmon is controlled by an external magnetic flux $\Phi_{\text{ext}}$ applied via mutual coupling to the $Z$ line. We bias the qubit to the maximum frequency where it is first-order insensitive to flux noise.
\paragraph{Device fabrication}

The transmons were fabricated by Qolab on a high-resistivity Si wafer ($>\SI{10}{k\Omega-\text{cm}}$) cleaned with diluted HF to remove native oxide prior to sputter deposition of an aluminum base layer. The base layer was defined on a positive photoresist layer written using a Heidelberg DWL 66+ laser writer, then developed and etched using a TMAH-based developer. Dolan bridge Josephson junctions were fabricated using electron beam lithography (EBL), including a bilayer stack of EBL resist exposed on an Elionix 100 keV electron beam writer and developed with a dilute solvent mixture. The counterelectrode was deposited using an electron beam evaporator. Liftoff was performed in an NMP-based solvent, then the sample was sonicated in a solvent bath before being diced and packaged.

\subsection{Experimental setup at MIT and Device B}
The measurements are performed in a Bluefors XLD-1000 dilution refrigerator with a base temperature of $\approx\SI{13}{\milli\kelvin}$. The anharmonicity of the qubit is approximately $\SI{-210}{\mega\hertz}$. The Quantum Machines OPX1000 is used for the XY and Z control of the qubit and readout signal, and both microwave pulses are generated by single-sideband modulation with suppressed carrier. For the qubit coherence experiment in the main text, each drive pulse is 20-ns long. The Z control lines are attenuated by 20 dB at the 4K stage of the cryostat, then filtered using a Minicircuits VLFX-300 low-pass filter and a Quantum Microwave QMC-CRYOIRF-001 low-cutoff IR filter at the mixing chamber. The XY control lines and readout tone, approximately $\SI{8.3}{\giga\hertz}$, are passively attenuated by 70 dB inside the cryostat and filtered with a Quantum Microwave QMC-CRYOIRF-004 high-cutoff IR filter. The device is shielded magnetically with a superconducting can. The transmitted signal from the feedline passes through a Quantum Microwave QMC-CRYOIRF-003 high-cutoff IR filter and a Low Noise Factory 4-8 GHz double-junction isolator to remove noise of the Josephson traveling wave parametric amplifier (JTWPA) and dump it in a $\SI{50}{\ohm}$ terminator. A Holzworth HS9000 Synthesizer pumps the JTWPA. The signal then passes through an RLC Electronics F-19704 high-pass filter and a Low Noise Factory 4-8 GHz single-junction isolator, thermally anchored at the mixing chamber stage, to prevent back reflections from the amplification chain. A high-electron-mobility-transistor amplifier (LNC0.3\_14B) thermally anchored at the $\SI{3}{\kelvin}$ stage amplifies the readout signal. At room temperature, the readout line is again amplified (Narda-MITEQ LNA-40-00101200-17-10P).

\paragraph{Device fabrication}
Device B was fabricated by MIT Lincoln Laboratory using a \SI{200}{\milli\meter} aluminum-on-silicon fabrication process. The Josephson junctions for the SQUID transmon were deposited by shadow evaporation using a Ge-based mask patterned for Dolan-bridge-style Al-AlOx-Al junctions~\cite{gingras2026}.

\newpage
\section{Telegraphic switching analysis}

\begin{figure*}[t]
    \centering
    \includegraphics{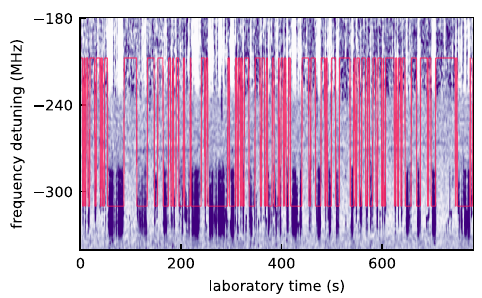}
    \caption{\textbf{Telegraphic switching analysis of a TLS feature (Device A).}
    Dynamic relaxation-rate map from the dataset used in Fig.~2(a) of the main text. The red curve shows the two-state trajectory extracted from two preselected telegraphic frequency bands.}
    \label{fig:SuppFigTelegraphic}
\end{figure*}

\begin{figure*}[t]
    \centering
    \includegraphics[width=0.52\textwidth]{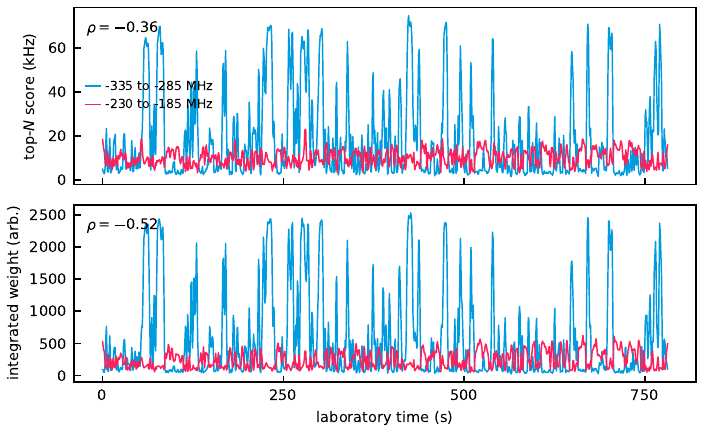}
    \caption{\textbf{Raw top-$N$ band scores and integrated band weights for the two frequency bands used in Fig.~\ref{fig:SuppFigTelegraphic} (Device A).}
    The two bands are anticorrelated, supporting transfer of spectral weight between the assigned states, although the integrated weight is not exactly conserved.}
    \label{fig:SuppFigTelegraphicRobustness}
\end{figure*}

We analyze the telegraphic TLS dynamics in Fig.~\ref{fig:SuppFigTelegraphic} using the same dataset as Fig.~2(a) of the main text. As in the spectral-diffusion analysis below, the measured relaxation time is first converted to a relaxation rate,
\begin{equation}
    \Gamma_1(f_i,t_n)=\frac{1}{T_1(f_i,t_n)}.
\end{equation}
The map is smoothed with a Gaussian kernel of width one pixel along the time and frequency axes. We then subtract the static frequency-dependent background,
\begin{equation}
    \Gamma_1^{\mathrm{dyn}}(f_i,t_n)
    =
    \Gamma_1^{\mathrm{sm}}(f_i,t_n)
    -
    \operatorname{median}_{n}\!\left[
    \Gamma_1^{\mathrm{sm}}(f_i,t_n)
    \right],
    \label{eq:telegraphic_dynamic_gamma}
\end{equation}
and restrict the analysis to the frequency interval from $\SI{-340}{\mega\hertz}$ to $\SI{-180}{\mega\hertz}$ containing the telegraphic feature.

The switching trajectory is extracted as a two-state process. We define two frequency bands corresponding to the observed TLS states,
\begin{equation}
    B_0=[\SI{-335}{\mega\hertz},\SI{-285}{\mega\hertz}],
    \qquad
    B_1=[\SI{-230}{\mega\hertz},\SI{-185}{\mega\hertz}].
\end{equation}
For each laboratory time $t_n$ and band $B_k$, we compute a band score $x_k(t_n)$ from the mean of the ten largest values of the positive dynamic signal inside the band,
\begin{equation}
    x_k(t_n)
    =
    \frac{1}{N_{\mathrm{top}}}
    \sum_{j\in \mathrm{top}(B_k,N_{\mathrm{top}})}
    \max\!\left[
    \Gamma_1^{\mathrm{dyn}}(f_j,t_n)-q_{10}(t_n),0
    \right],
    \label{eq:telegraphic_band_score}
\end{equation}
where $N_{\mathrm{top}}=10$ and $q_{10}(t_n)$ is the 10th percentile of $\Gamma_1^{\mathrm{dyn}}(f,t_n)$ within the analyzed frequency interval. From these scores we extract the switching rate in three steps: a hysteresis decoding of the two-state trajectory, a hidden-Markov analysis that propagates the state-assignment uncertainty, and an Allan-deviation cross-check of the resulting rates.

\subsection{Hysteresis filtering}
We now assign the observed signal $\mathbf{x}_n=[{x}_0(t_n),{x}_1(t_n)]$ to a hidden state of two-level fluctuator $s_n\in \{0,1\}$. The simplest assignment applies a hysteresis rule to the score difference: a state switch is accepted only when the difference between the signals exceeds its own scatter. In particular, we define the score difference as 
\begin{equation}
    \Delta \overline x(t_n)=\overline{x}_1(t_n)-\overline{x}_0(t_n),
\end{equation}
where we additionally smooth each score over a short window, $\overline{x}_k(t_n)=\tfrac{1}{w}\sum...$, with $w=3$ samples ($\approx\SI{2.29}{\second}$). Next we compare against the finite threshold $h = 0.1\times \text{std}(\Delta \overline{x})$. We assume the hidden state $s_n\in\{0,1\}$ switches from 0 to 1 only when $\Delta \overline x(t_n)>h$, and from 1 to 0 only when $\Delta \overline x(t_n)<-h$. The plotted trajectory is then
\begin{equation}
    f_{\mathrm{TLS}}(t_n)
    =
    \frac{1}{2}
    \left[
    f_{\min}(B_{s_n})+f_{\max}(B_{s_n})
    \right].
    \label{eq:telegraphic_two_state_trajectory}
\end{equation}
Thus the red curve in Fig.~\ref{fig:SuppFigTelegraphic} reports the assigned discrete state. The raw band scores have a correlation coefficient $\rho=-0.36$, strengthening to $\rho=-0.52$ when the positive signal is integrated over all frequencies in each band (Fig.~\ref{fig:SuppFigTelegraphicRobustness}). This anticorrelation supports the interpretation of spectral-weight transfer between the two assigned frequency states, but it does not by itself prove exact conservation of the integrated loss amplitude.

Counting transitions directly on this decoded trajectory gives $106$ switches, state populations $(0.52,0.48)$, and rates $k_{01}=\SI{0.134}{\hertz}$, $k_{10}=\SI{0.144}{\hertz}$, so that $1/(k_{01}+k_{10})=\SI{3.59}{\second}$. This hard-threshold count returns a single value with no uncertainty and is biased toward long dwell times: the score smoothing merges rapid back-and-forth switches below the temporal resolution, so many short dwells are likely omitted from the count and the inferred rate is underestimated. Establishing the switching rate therefore requires a model-based analysis that propagates the state-assignment ambiguity rather than committing to a single decoded path, which we develop next.

\subsection{Bayesian estimation of the switching rates}
\begin{figure*}[t]
    \centering
    \includegraphics[width=\textwidth]{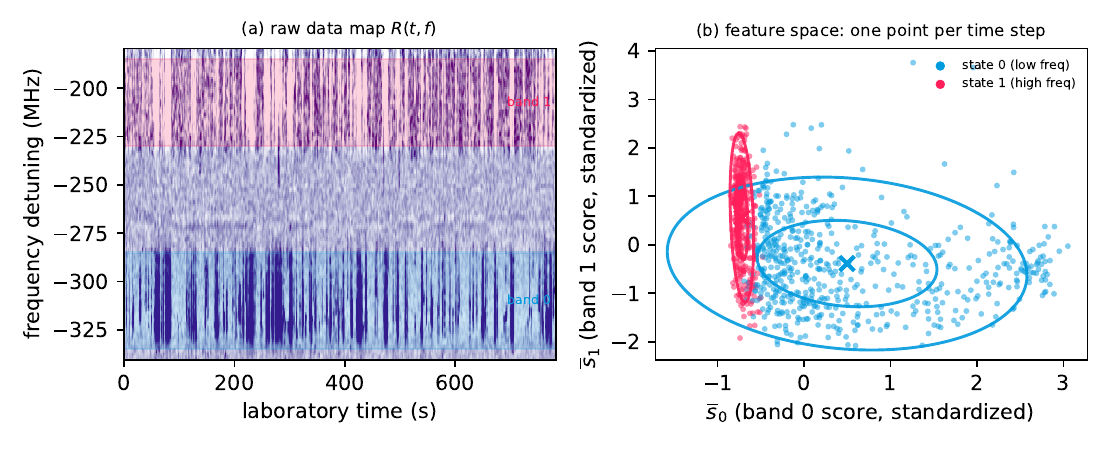}
    \caption{\textbf{Gaussian emission model (Device A).}
    (a) The raw data is the dynamic relaxation-rate map. (b) Each time step is reduced to the two band scores $(\overline{x}_0,\overline{x}_1)$, giving one point per time step in the standardized score plane. The points form two clusters, and the two-state Gaussian HMM fits one full-covariance Gaussian $(\boldsymbol{\mu}_k,\Sigma_k)$ per cluster ($1\sigma$ and $2\sigma$ ellipses; $\times$ marks $\boldsymbol{\mu}_k$). Points are colored by the Viterbi state.}
    \label{fig:SuppFigEmission}
\end{figure*}

The above analysis establishes telegraphic switching but returns a single decoded trajectory, and hence a single value per rate with no statistical uncertainty. We therefore add a hidden Markov model (HMM)~\cite{Rabiner1989} that propagates two sources of uncertainty: the ambiguity of the state assignment at each time step, and the finite-sample noise of counting a limited number of switches. By assigning probabilities to paths rather than committing to a single decoding, it recovers the short dwells that the hysteresis count merges and returns a full posterior for each rate. The hidden state is the TLS state $s_n\in\{0,1\}$, and the observation is the band-score vector $\mathbf{x}_n=[\overline{x}_0(t_n),\overline{x}_1(t_n)]$ of Eq.~\eqref{eq:telegraphic_band_score}, standardized per component. Conditioned on the state, the observation is a full-covariance Gaussian,
\begin{equation}
    p(\mathbf{x}_n\mid s_n=k)=\mathcal{N}(\mathbf{x}_n;\boldsymbol{\mu}_k,\Sigma_k),
    \label{eq:telegraphic_hmm_emission}
\end{equation}
so each state is a Gaussian cluster in the two-dimensional score plane (Fig.~\ref{fig:SuppFigEmission}). The hidden state evolves through a discrete transition matrix at spacing $\Delta t$,
\begin{equation}
    \Afit(i,j)=P(s_{n+1}=j\mid s_n=i),
    \qquad
    i,j\in\{0,1\}.
    \label{eq:telegraphic_hmm_transmat}
\end{equation}
The emission parameters $\{\boldsymbol{\mu}_k,\Sigma_k\}$ and $\Afit$ are estimated \emph{jointly} from the sequence $\{\mathbf{x}_n\}$ by Baum--Welch expectation maximization~\cite{BaumWelch1970}, which maximizes the data log-likelihood $\mathcal{L}=\log\sum_i\alpha_T(i)$ with $\alpha$ the forward quantity of Eq.~\eqref{eq:ffbs_forward}; we use full covariances and $500$ EM iterations. This replaces the fixed threshold and hysteresis rule of Eq.~\eqref{eq:telegraphic_two_state_trajectory} by a learned, probabilistic state assignment. For this dataset the two fitted clusters are well separated (Fig.~\ref{fig:SuppFigEmission}), and the most probable trajectory (Viterbi algorithm) of the fitted model agrees with the hysteresis trajectory at the $97\%$ level.

\subsubsection{Forward-filtering backward-sampling}
\begin{figure}[t]
    \centering
    \begin{tikzpicture}[
        st/.style={circle, draw, minimum size=6mm, font=\footnotesize, inner sep=0pt},
        dim/.style={circle, draw=gray!40, minimum size=6mm, inner sep=0pt},
        lbl/.style={font=\footnotesize},
        sel/.style={-{Stealth[length=1.8mm]}, very thick, cRed},
        e/.style={-{Stealth[length=1.3mm]}, gray!45},
    ]
    \begin{scope}
    \foreach \c in {1,...,5}{
        \pgfmathsetmacro\x{1.15*\c}
        \pgfmathsetmacro\sa{25+55*abs(sin(\c*57))}
        \pgfmathsetmacro\sb{25+55*abs(cos(\c*41))}
        \node[st, fill=cPurple!\sa] (a\c) at (\x,1.0) {};
        \node[st, fill=cPurple!\sb] (b\c) at (\x,0)   {};
        \node[font=\scriptsize, below=0.5mm of b\c] {$t_{\c}$};
    }
    \node[lbl, left=1.5mm of a1] {$s{=}1$};
    \node[lbl, left=1.5mm of b1] {$s{=}0$};
    \foreach \c [evaluate=\c as \n using int(\c+1)] in {1,...,4}{
        \draw[e] (a\c) -- (a\n); \draw[e] (a\c) -- (b\n);
        \draw[e] (b\c) -- (a\n); \draw[e] (b\c) -- (b\n);
    }
    \node[lbl] at (3.45,1.9) {(a) forward: fill all $\alpha_n(i)$};
    \end{scope}
    \begin{scope}[xshift=7.4cm]
    \def\pth{{0,0,1,1,0}}
    \foreach \c in {1,...,5}{
        \pgfmathsetmacro\x{1.15*\c}
        \pgfmathtruncatemacro\sel{\pth[\c-1]}
        \ifnum\sel=1
            \node[st, fill=cRed!70, text=white] (c\c) at (\x,1.0) {\c};
            \node[dim] (d\c) at (\x,0) {};
        \else
            \node[dim] (c\c) at (\x,1.0) {};
            \node[st, fill=cRed!70, text=white] (d\c) at (\x,0) {\c};
        \fi
        \node[font=\scriptsize, below=0.5mm of d\c] {$t_{\c}$};
    }
    \node[lbl, left=1.5mm of c1] {$s{=}1$};
    \node[lbl, left=1.5mm of d1] {$s{=}0$};
    \node (q1) at (d1) {}; \node (q2) at (d2) {}; \node (q3) at (c3) {};
    \node (q4) at (c4) {}; \node (q5) at (d5) {};
    \draw[sel] (q5) -- (q4); \draw[sel] (q4) -- (q3);
    \draw[sel] (q3) -- (q2); \draw[sel] (q2) -- (q1);
    \node[lbl] at (3.45,1.9) {(b) backward: sample one path};
    \end{scope}
    \end{tikzpicture}
    \caption{\textbf{Forward-filtering backward-sampling.}
    (a) The forward pass fills every lattice node with the filtered weight $\alpha_n(i)$ of Eq.~\eqref{eq:ffbs_forward} (shading $\propto\alpha$); no randomness. (b) One path is then sampled from right to left via Eq.~\eqref{eq:ffbs_sample}; the factor $\Afit(i,s_{n+1})$ forces each state to connect legally to the one already chosen at $t_{n+1}$. Repeating (b) gives $N_{\mathrm{p}}=400$ posterior paths.}
    \label{fig:SuppFigTrellis}
\end{figure}
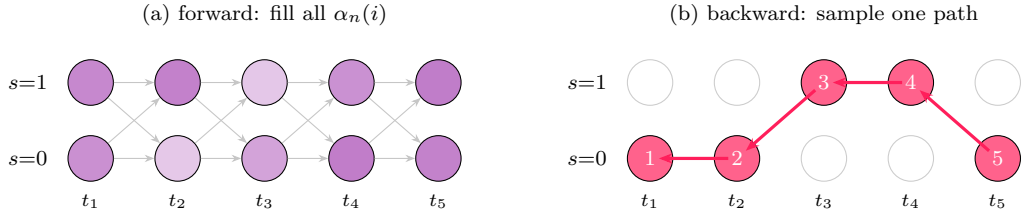

To propagate the state-assignment uncertainty we draw $N_{\mathrm{p}}=400$ full state paths from the exact posterior $P(s_{1:T}\mid\mathbf{x}_{1:T})$ by forward-filtering backward-sampling~\cite{CarterKohn1994,FruhwirthSchnatter1994}. The forward pass computes the filtered joint likelihood, and the backward pass the complementary future likelihood,
\begin{align}
    \alpha_n(i)&=p(\mathbf{x}_{1:n},s_n=i)
    =p(\mathbf{x}_n\mid s_n=i)\sum_j\alpha_{n-1}(j)\,\Afit(j,i),
    \label{eq:ffbs_forward}\\
    \beta_n(i)&=p(\mathbf{x}_{n+1:T}\mid s_n=i)
    =\sum_j \Afit(i,j)\,p(\mathbf{x}_{n+1}\mid s_{n+1}=j)\,\beta_{n+1}(j),
    \label{eq:ffbs_backward}
\end{align}
both evaluated in log space for numerical stability and both using the same Gaussian emission of Eq.~\eqref{eq:telegraphic_hmm_emission} (Fig.~\ref{fig:SuppFigTrellis}). A path is then drawn from right to left,
\begin{equation}
    s_T\sim\alpha_T(i),
    \qquad
    s_n\sim\alpha_n(i)\,\Afit(i,s_{n+1}),
    \qquad n=T-1,\dots,1,
    \label{eq:ffbs_sample}
\end{equation}
so every draw is temporally consistent and conditions on the full record. The draws agree where the signal is clean and fan out where the jump timing is ambiguous.

\subsubsection{Counting uncertainty}
\begin{figure*}[t]
    \centering
    \includegraphics[width=\textwidth]{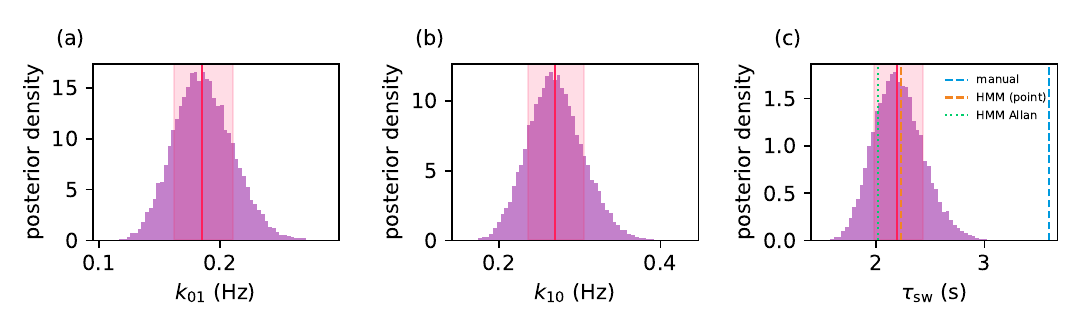}
    \caption{\textbf{Posterior distributions of the switching rates (Device A).}
    Posterior histograms for (a) $k_{01}$, (b) $k_{10}$, and (c) the characteristic switching time $\tau_{\mathrm{sw}}=1/(k_{01}+k_{10})$, from the $2\times10^4$ pooled samples of Eq.~\eqref{eq:bayes_rates}. The solid line marks the median and the shaded band the $68\%$ credible interval. Panel (c) overlays the manual-hysteresis and hidden-Markov point estimates for comparison.}
    \label{fig:SuppFigBayesRates}
\end{figure*}

Given a sampled path the two states are fixed, but the rates are still only finitely sampled: we observe a limited number of switches, and this counting noise must be propagated. Conditional on a path, each time step in state $0$ independently transitions to state $1$ with probability $p_{01}$, so the number of such transitions among the $n_0$ time steps in state $0$ is binomial, $N_{01}\mid s_{1:T}\sim\mathrm{Binomial}(n_0,p_{01})$. With a conjugate $\mathrm{Beta}(a,b)$ prior the transition probability therefore has the exact posterior \begin{equation} p_{01}\mid s_{1:T}\;\sim\;\mathrm{Beta}(a+N_{01},\,b+n_0-N_{01}), \label{eq:beta_posterior} \end{equation} and analogously for $p_{10}$; we use a uniform prior $a=b=1$. The two-state chain has eigenvalues $1$ and $\lambda_2=1-p_{01}-p_{10}$; matching this eigenvalue and the stationary distribution to those of the continuous-time generator gives the closed-form rates
\begin{equation}
    \gamma=k_{01}+k_{10}=-\frac{\ln\lambda_2}{\Delta t},
    \qquad
    k_{01}=\gamma\,\frac{p_{01}}{p_{01}+p_{10}},
    \qquad
    k_{10}=\gamma\,\frac{p_{10}}{p_{01}+p_{10}},
    \label{eq:rate_map}
\end{equation}
valid while $\lambda_2>0$. Drawing $(p_{01},p_{10})$ from Eq.~\eqref{eq:beta_posterior} and replacing them in Eq.~\eqref{eq:rate_map} yields posterior samples of $(k_{01},k_{10},\tau_{\mathrm{sw}}=1/\gamma)$. Pooling $50$ parameter draws over each of the $400$ sampled paths ($2\times10^4$ samples) marginalizes both sources of uncertainty at once, giving (median with $68\%$ credible interval):
\begin{align}
    k_{01}&=0.185^{+0.026}_{-0.023}\,\si{\hertz}, &
    k_{10}&=0.269^{+0.036}_{-0.032}\,\si{\hertz}, &
    \tau_{\mathrm{sw}}&=2.20^{+0.24}_{-0.22}\,\si{\second}.
    \label{eq:bayes_rates}
\end{align}
The posterior distributions are shown in Fig.~\ref{fig:SuppFigBayesRates}. The direct-counting value $\SI{3.59}{\second}$ of Sec.~II A lies outside the
$95\%$ credible interval $[\SI{1.80}{\second},\SI{2.70}{\second}]$ of the
posterior, consistent with the merging of short dwells by the hysteresis
smoothing.

\subsection{Allan deviation}
\begin{figure}[t]
    \centering
    \includegraphics[width=0.52\textwidth]{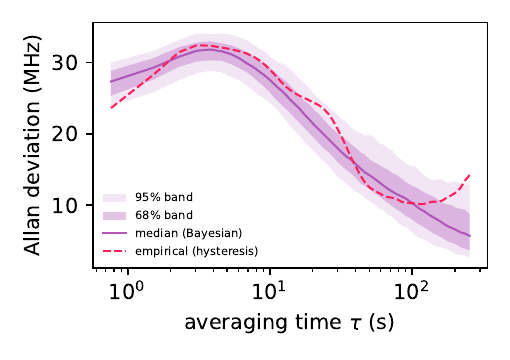}
    \caption{\textbf{Allan-deviation credible band from the rate posterior (Device A).}
    Median (solid) and $68\%$/$95\%$ credible bands of the Allan deviation, obtained by simulating a two-state trajectory from each posterior $(k_{01},k_{10})$ draw through $A=\exp(Q\,\Delta t)$ and computing its overlapping Allan deviation. The dashed curve is the empirical Allan deviation of the extracted trajectory.}
    \label{fig:SuppFigBayesAllan}
\end{figure}

As an independent check of the hidden-Markov rates that does not rely on a spectral fit, we examine the overlapping Allan deviation computed from the single trajectory and from trajectories sampled from the rate posterior. For an averaging time $\tau_m = m\,\Delta t$, the discrete moving average is
\begin{equation}
\overline f_{n,m} = \frac{1}{m}\sum_{\ell=0}^{m-1} f_{\mathrm{TLS}}(t_{n+\ell}),
\label{eq:telegraphic_movavg}
\end{equation}
and the overlapping Allan deviation is
\begin{equation}
\sigma_f(\tau_m) = \sqrt{\tfrac{1}{2}\big\langle \big(\overline f_{n+m,m}-\overline f_{n,m}\big)^2\big\rangle_n }.
\label{eq:telegraphic_allan}
\end{equation}
A two-state Markov (telegraphic) process has a single correlation time, and hence a single Lorentzian spectrum; its Allan deviation therefore rises as $\sqrt{\tau}$ at short averaging times, peaks near the switching time, and falls as $1/\sqrt{\tau}$ afterward. The measured $\sigma_f(\tau)$ shows this single-humped shape (Fig.~\ref{fig:SuppFigBayesAllan}), indicating a single few-second timescale rather than a distribution of rates or a $1/f$ background.
 
We attach uncertainty to the curve directly from the rate posterior. For each of $300$ posterior draws $(k_{01},k_{10})$ [Eq.~\eqref{eq:bayes_rates}] we simulate a two-state trajectory of the same length using the discrete transition matrix $A=\exp(Q\,\Delta t)$, with generator $Q=\left(\begin{smallmatrix}-k_{01}&k_{01}\\ k_{10}&-k_{10}\end{smallmatrix}\right)$ and levels fixed at the band centers, and compute its overlapping Allan deviation [Eq.~\eqref{eq:telegraphic_allan}]. Percentiles across the $300$ curves give a credible band that is consistent by construction with the rate posterior (Fig.~\ref{fig:SuppFigBayesAllan}). The empirical Allan deviation of the extracted trajectory lies within this band up to $\tau\simeq\SI{40}{\second}$. The upturn beyond is the expected finite-sample estimator artifact at large lag rather than a physical feature. The peak of $\sigma_f(\tau)$ falls near four seconds. Since the overlapping Allan deviation of a telegraphic process peaks at $\tau_\text{peak}\approx1.9\,\tau_{\mathrm{sw}}$, this corresponds to $\tau_{\mathrm{sw}}\approx\SI{2}{\second}$, confirming the hidden-Markov estimate $\tau_{\mathrm{sw}}=\SI{2.20}{\second}$ (Eq.~\eqref{eq:bayes_rates}) and, in particular, the shorter switching time recovered once the merged short dwells are accounted for. 

\newpage

\section{Spectral diffusion analysis}

\begin{figure*}[t]
    \centering
    \includegraphics{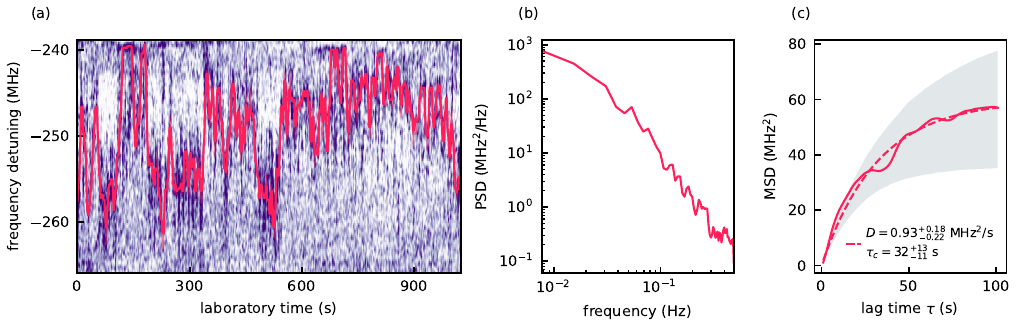}
    \caption{\textbf{Spectral diffusion analysis of a fluctuating TLS (Device A).}
    (a) Dynamic relaxation-rate map from the dataset used in Fig.~2(b) after subtracting the static frequency-dependent background. The red curve highlights the tracked trajectory of the first TLS feature.
    (b) Power spectral density of the tracked TLS frequency trajectory.
    (c) Mean-squared displacement of the tracked trajectory with a fit to an Ornstein-Uhlenbeck diffusion model. The shaded band is the 68\% interval obtained by block bootstrapping the tracked trajectory.}
    \label{fig:SuppFigDiffusive}
\end{figure*}

To quantify the spectral diffusion of the fluctuating TLS highlighted in Fig.~\ref{fig:SuppFigDiffusive}(a), same as Fig.2(b) of the main text, we first convert the measured relaxation time to a relaxation rate,
\begin{equation}
    \Gamma_1(f_i,t_n)=\frac{1}{T_1(f_i,t_n)}.
\end{equation}
We then smooth the map with a Gaussian kernel of width one pixel along both the time and frequency axes, and subtract the time-independent background at each frequency point,
\begin{equation}
    \Gamma_1^{\mathrm{dyn}}(f_i,t_n)
    =
    \Gamma_1^{\mathrm{sm}}(f_i,t_n)
    -
    \operatorname{median}_{n}\!\left[
    \Gamma_1^{\mathrm{sm}}(f_i,t_n)
    \right].
    \label{eq:dynamic_gamma}
\end{equation}
This subtraction suppresses static horizontal features in the spectroscopy map and leaves the time-dependent component of the loss feature.

The TLS trajectory is extracted from a frequency window around the feature using a Viterbi tracker. Let $S_{n,j}=\Gamma_1^{\mathrm{dyn}}(f_j,t_n)$ be the dynamic signal in this window. The tracked frequency-index sequence $\{j_n\}$ is chosen to maximize
\begin{equation}
    \mathcal{L}(\{j_n\})
    =
    \sum_{n=1}^{N_t} S_{n,j_n}
    -
    \lambda
    \sum_{n=2}^{N_t} |j_n-j_{n-1}| ,
    \label{eq:viterbi_tls}
\end{equation}
with a jump penalty $\lambda=10$. The resulting frequency trajectory is $f_{\mathrm{TLS}}(t_n)=f_{j_n}$.

We analyze the trajectory using both its power spectral density and mean-squared displacement. For the power spectral density shown in Fig.~\ref{fig:SuppFigDiffusive}(b), we subtract the mean frequency and estimate
\begin{equation}
    S_f(\nu)
    =
    \frac{1}{T}
    \left|
    \int_0^T
    \left[
    f_{\mathrm{TLS}}(t)-\langle f_{\mathrm{TLS}}\rangle
    \right]
    e^{-i2\pi \nu t}\,dt
    \right|^2 ,
    \label{eq:psd_tls}
\end{equation}
using Welch averaging in the numerical implementation.

The mean-squared displacement is computed from the discrete trajectory as
\begin{equation}
    \mathrm{MSD}(\tau_m)
    =
    \frac{1}{N_t-m}
    \sum_{n=1}^{N_t-m}
    \left[
    f_{\mathrm{TLS}}(t_{n+m})-f_{\mathrm{TLS}}(t_n)
    \right]^2 ,
    \qquad
    \tau_m=m\Delta t .
    \label{eq:msd_tls}
\end{equation}
We fit this quantity to the Ornstein-Uhlenbeck form
\begin{equation}
    \mathrm{MSD}(\tau)
    =
    \frac{2D}{\gamma}
    \left(1-e^{-\gamma \tau}\right),
    \qquad
    \tau_c=\frac{1}{\gamma},
    \label{eq:ou_msd_tls}
\end{equation}
The nominal analysis yields
\begin{equation}
    D
    =
    0.93^{+0.18}_{-0.22}\,
    \si{\mega\hertz\squared\per\second},
    \qquad
    \tau_c
    =
    32^{+13}_{-11}\,
    \si{\second},
    \label{eq:ou_fit_results}
\end{equation}
where the uncertainties are 68\% block-bootstrap intervals. Because the MSD values at neighboring lags are correlated, we do not use the covariance matrix of the nonlinear least-squares fit as the uncertainty estimate. Instead, we draw five contiguous blocks of 200 samples (\(\approx\SI{204}{\second}\)) with replacement from the tracked trajectory. The squared displacements entering Eq.~\eqref{eq:msd_tls} are evaluated only within each block, avoiding artificial jumps at block boundaries, and the OU model is refitted for each of 1000 bootstrap realizations. The resulting 95\% intervals are \(D\in[0.57,1.28]\,\si{\mega\hertz\squared\per\second}\) and \(\tau_c\in[13,61]\,\si{\second}\).

We additionally rerun the complete smoothing, background-subtraction, Viterbi-tracking, and fitting procedure over a \(3\times3\times3\) robustness grid. The frequency-window width is varied between \(0.8\), \(1.0\), and \(1.2\) times its nominal value; the Gaussian smoothing standard deviation is varied between \(0.5\), \(1.0\), and \(1.5\) samples along both map axes; and the Viterbi jump penalty is varied between \(5\), \(10\), and \(20\). Across all 27 combinations, the fitted values remain within
\begin{equation}
    D\in[0.70,0.96]\,
    \si{\mega\hertz\squared\per\second},
    \qquad
    \tau_c\in[23,38]\,
    \si{\second}.
\end{equation}
When varied one parameter at a time, the corresponding ranges are \(D=0.76\)--\(0.93\,\si{\mega\hertz\squared\per\second}\) and \(\tau_c=26\)--\(\SI{32}{\second}\) for the ROI width, \(D=0.89\)--\(0.94\,\si{\mega\hertz\squared\per\second}\) and \(\tau_c=29\)--\(\SI{38}{\second}\) for smoothing, and \(D=0.90\)--\(0.93\,\si{\mega\hertz\squared\per\second}\) and \(\tau_c=32\)--\(\SI{37}{\second}\) for the Viterbi penalty. Thus, the conclusion that this feature is a rapidly diffusing TLS is insensitive to reasonable analysis choices, while the bootstrap interval captures the larger uncertainty associated with finite-duration sampling of the stochastic trajectory.

\newpage

\section{On- and off-TLS randomized benchmarking comparison}

\begin{figure*}[t]
    \centering
    \includegraphics{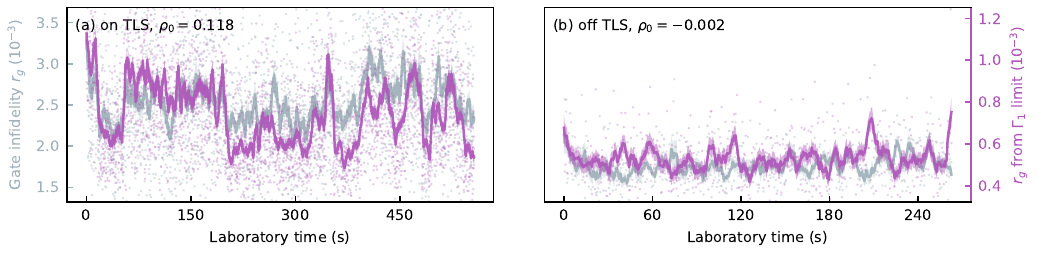}
    \caption{\textbf{Comparison of randomized-benchmarking infidelity on and off the TLS feature (Device B).}
    (a) Gate infidelity extracted from randomized benchmarking and the corresponding relaxation-limited contribution for the experiment of Fig.~3(b) of the main text.
    (b) Same experiment, acquired away from the TLS feature. Dots show the raw single-repetition values, curves are 5-s moving averages, and shaded regions show 68\% confidence intervals. The off-TLS measurement exhibits a zero-lag correlation consistent with zero and a lower average relaxation-limited contribution than the on-TLS measurement.}
    \label{fig:SuppFigOnOffTLS}
\end{figure*}

Figure~\ref{fig:SuppFigOnOffTLS} compares the randomized-benchmarking gate infidelity $r_g$ to the infidelity expected from energy relaxation for two otherwise similar interleaved measurements. The points are the raw values from individual repetitions, while the solid curves are moving averages. Panel (a) shows the on-TLS experiment of Fig.~3(b) of the main text, while in (b) the same interleaved experiment is performed a few MHz away from the TLS feature of (a). For each repetition, the relaxation-limited contribution is computed as in Ref.~\cite{Malley2015}:
\begin{equation}
    r_{\Gamma_1}(t_n)
    =
    \frac{t_{\mathrm{gate}}}{3T_1(t_n)},
    \label{eq:rgamma1_limit}
\end{equation}
with $t_{\mathrm{gate}}=\SI{20}{\nano\second}$. Both traces are plotted as moving averages over a centered time window of width $\SI{5}{\second}$,
\begin{equation}
    \overline{x}(t_n)
    =
    \frac{1}{N_n}
    \sum_{m:\, |t_m-t_n|\leq \SI{2.5}{\second}}
    x(t_m).
    \label{eq:onoff_moving_average}
\end{equation}
For the RB trace, the shaded interval is the standard error of the values inside the same moving window. For the relaxation-limited trace, the shaded interval is obtained by propagating the posterior standard deviation of $T_1$ through Eq.~\eqref{eq:rgamma1_limit} and then computing the standard error of the windowed mean,
\begin{equation}
    \sigma_{\overline{r}_{\Gamma_1}}(t_n)
    =
    \frac{1}{N_n}
    \sqrt{
    \sum_{m:\, |t_m-t_n|\leq \SI{2.5}{\second}}
    \left[
    \frac{t_{\mathrm{gate}}}{3T_1(t_m)^2}\,
    \sigma_{T_1}(t_m)
    \right]^2
    }.
    \label{eq:onoff_t1_limit_sem}
\end{equation}

To quantify whether the measured RB infidelity follows the relaxation-limited trace, we compute the zero-lag normalized cross-correlation. Let $x_n=r_g(t_n)$ and $y_n=r_{\Gamma_1}(t_n)$. The correlation is
\begin{equation}
    \rho_0
    =
    \frac{
    \sum_n
    \left(x_n-\overline{x}\right)
    \left(y_n-\overline{y}\right)
    }
    {
    \sqrt{
    \sum_n \left(x_n-\overline{x}\right)^2
    \sum_n \left(y_n-\overline{y}\right)^2
    }}
    ,
    \label{eq:onoff_cross_correlation}
\end{equation}
Statistical significance is estimated using a circular-shift permutation test: the $y_n$ trace is randomly rolled by a nonzero number of samples, and the zero-lag correlation is recomputed. The empirical p-value is
\begin{equation}
    p
    =
    \frac{1+N_{\mathrm{null}}\left(|\rho_{\mathrm{null}}|\geq |\rho_{\mathrm{obs}}|\right)}
    {1+N_{\mathrm{perm}}},
    \label{eq:onoff_pvalue}
\end{equation}
with $N_{\mathrm{perm}}=2000$.

For the same $\SI{5}{\second}$ moving-averaged traces shown in Fig.~\ref{fig:SuppFigOnOffTLS}, the on-TLS dataset has a zero-lag correlation of $\rho_0=0.618$ with $p=5.0\times10^{-4}$. For the off-TLS dataset, the corresponding correlation is $\rho_0=-0.0507$ with $p=0.64$. Thus, the on-TLS measurement exhibits a strong and statistically significant correlation between the RB infidelity and the relaxation-limited contribution, whereas the off-TLS control is consistent with no correlation. In addition, the off-TLS measurement exhibits both a lower mean relaxation rate and a lower mean RB infidelity than the on-TLS measurement.
\newpage

\section{Wide frequency scans}

\begin{figure*}[t]
    \centering
    \includegraphics[width=\textwidth]{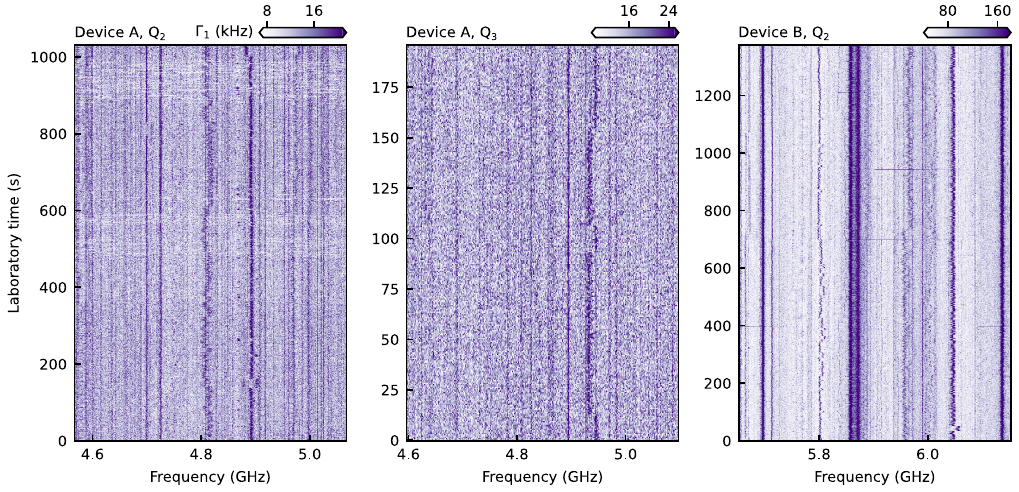}
    \caption{\textbf{Wide frequency scans for the devices used in the main text.}
    Relaxation-rate maps over wider frequency intervals for two qubits on Device A and the qubit of Device B.}
    \label{fig:SuppFigWideScan}
\end{figure*}

Figure~\ref{fig:SuppFigWideScan} shows wide frequency scans acquired on the devices used in the main text. The first two panels correspond to Device A and show the wider spectral environment of Q$_2$ and Q$_3$, the third panel corresponds to Q$_2$ of Device B from a following cooldown. The maps show that the zoomed features analyzed in the main text are embedded in a broader and dense spectrum of relaxation-rate features, consistent with coupling to multiple TLS defects across the tunable qubit bandwidth.

\bibliography{my_bibliography}